\documentstyle[pra,aps,twocolumn,epsfig]{revtex}
\begin{document}
\draft
\title{Dielectronic recombination of lithium-like Ni$^{25+}$ ions --- high resolution rate coefficients and
influence of external crossed E and B fields}

\author{S. Schippers, T. Bartsch, C. Brandau,  A. M\"uller}
\address{Institut f\"ur Kernphysik, Universit\"at Giessen, 35392
Giessen, Germany}

\author{G. Gwinner, G. Wissler, M. Beutelspacher, M. Grieser, A. Wolf}
\address{Max-Planck-Institut f\"ur Kernphysik and
  Physikalisches Institut der Universit\"at Heidelberg, 69117
  Heidelberg, Germany}

\author{R. A. Phaneuf}
\address{Department of Physics, University of  Nevada, Reno, NV 89557, USA}

\date{\today}
\maketitle

\begin{abstract}

Absolute dielectronic recombination (DR) rates for lithium-like
Ni$^{25+}$($1s^2 2s$) ions were measured at high-energy resolution
at the Heidelberg heavy-ion storage ring TSR. We studied the
center-of-mass energy range 0--130 eV which covers all $\Delta
n$=0 core excitations. The influence of external crossed electric
(0--300 V/cm) and magnetic (41.8--80.1 mT) fields was
investigated. For the measurement at near-zero electric field
resonance energies and strengths are given for Rydberg levels up
to n$=$32; also Maxwellian plasma rate coefficients for the
$\Delta n$=0 DR at electron temperatures between 0.5 and 200 eV
are provided. For increasing electric field strength we find that
for both the $2p_{1/2}$ and the $2p_{3/2}$ series of
Ni$^{24+}$($1s^2 2p_j n\ell$) Ryd\-berg resonances with $n>30$ the
DR rate coefficient increases approximately linearly by up to a
factor of 1.5. The relative increase due to the applied electric
field for Ni$^{25+}$ is remarkably lower than that found in
previous measurements with lighter isoelectronic Si$^{11+}$,
Cl$^{14+}$ and also Ti$^{19+}$ ions, [T. Bartsch et al, Phys.\
Rev.\ Lett.\ {\bf 79}, 2233 (1997); {\bf 82}, 3779 (1999) and to
be published] and in contrast to the results for lighter ions no
clear dependence of the electric field enhancement on the magnetic
field strength is found. The Maxwellian plasma rate coefficients
for $\Delta n$=0 DR of Ni$^{25+}$ are enhanced by at most $11\%$
in the presence of the strongest experimentally applied fields.
\end{abstract}


\section{Introduction}

Dielectronic recombination (DR) is an electron-ion collision
process which is well known to be important in astrophysical and
fusion plasmas\cite{DV80,Mue95}. In DR the initially free electron
is transferred to a bound state of the ion via a doubly excited
intermediate state which is formed by an excitation of the core
and a simultaneous attachment of the incident electron.
This two step process
\begin{equation}
  e^- + A^{q+} \to [A^{(q-1)+}]^{**} \to A^{(q-1)+}+ h\nu
\label{eq:DR}
\end{equation}
involves dielectronic capture (time-inverse Auger process) as the first step with a
subsequent stabilization of the lowered charge state by radiative decay to a state below the
ionization limit. This second step competes with autoionization which would transfer the ion
back into its initial charge state $q$ with the net effect being resonant elastic or
inelastic electron scattering. Another recombination process, which in contrast to DR is
non-resonant, is radiative recombination (RR)
\begin{equation}
  e^- + A^{q+} \to A^{(q-1)+}+ h\nu
\label{eq:RR}
\end{equation}
where the initially free electron is transferred to a bound state of the ion and a photon is
emitted simultaneously. The cross section for RR diverges at zero electron energy and
decreases rapidly towards higher energies. In the present investigation we regard RR as a
continuous background on top of which DR resonances are emerging.

In the case of narrow non-overlapping DR resonances the DR cross section
due to an intermediate state labelled $d$ can be well approximated
by\cite{sho69}
\begin{equation}
\sigma_d(E_{\rm cm}) = \bar{\sigma}_d L_d(E_{\rm cm})
\end{equation}
with the electron-ion center-of-mass (c.~m.) frame energy $E_{\rm cm}$, the Lorentzian line
shape $L_d(E)$ normalized to $\int L_d(E) dE=1$ and the resonance strength
\begin{eqnarray}
  \bar{\sigma}_d &=&  4.95\times10^{-30}{\rm cm}^2{\rm eV}^2{\rm s}\nonumber \\
  &&\times\frac{1}{E_d}
   \frac{g_d}{2 g_i}
   \frac{A_{\rm a}(d\to i) \sum_f A_{\rm r}(d\to f)}
                   {\sum_k A_{\rm a}(d\to k) + \sum_{f'}A_{\rm r}(d\to f')}
                   \label{eq:sigma}
\end{eqnarray}
where $E_d$ is the resonance energy, $g_i$ and $g_d$ are the statistical
weights of the initial ionic core $i$ and the doubly excited intermediate
state $d$, $A_{\rm a}(d\to i)$ and $A_{\rm r}(d \to f)$ denote the rate
for an autoionizing transition from $d$ to $i$ and the rate for a radiative
transition from $d$ to states $f$ below the first ionization limit,
respectively. The summation indices $k$ and $f'$ run over all states which
from $d$ can either be reached by autoionization or by radiative
transitions, respectively.

Soon after the establishment of DR as an important process governing the charge state
balance of ions in the solar corona\cite{bur64}, Burgess and Summers\cite{bur69} and Jacobs
et al.\cite{jac76} realized that DR cross sections should be sensitive to external electric
fields present in virtually any plasma environment. The electric field enhancement of DR
rates was subsequently reproduced in a number of theoretical calculations\cite{Hahn97}.
Briefly, the effect arises from the Stark mixing of $\ell$ states and the resulting
influence on the autoionization rates which, by detailed balancing, determine the capture of
the free electron. Autoionization rates strongly decrease with increasing $\ell$ and,
therefore, only low $\ell$ states significantly contribute to DR. Electric fields mix low
and high $\ell$ states and thereby increase the autoionization rates of the high $\ell$
states and consequently also the contribution of high Rydberg states to DR.

The control of external fields in experiments using intense ion
and electron beams is a challenge. Results from early
recombination experiments\cite{DS92,And92} could only be brought
in agreement with theory under the assumption that external
electric fields had been present in the interaction region. The
first experiment where external fields were applied under well
controlled conditions was performed by M\"{u}ller and
coworkers\cite{Mueller86/87} who investigated DR in the presence
of external fields (DRF) of singly charged Mg$^+$ ions. They
observed an increase of the measured DR cross section by a factor
of about 1.5 when increasing the motional $\vec{v}\times\vec{B}$
electric field from 7.2 to 23.5 V/cm. The agreement of these
results with theoretical predictions\cite{LNH86,BGP86} was at the
20\% level. Further DRF experiments with multiply charged C$^{3+}$
ions also revealed drastic DR rate enhancements by electric
fields\cite{Young94}; however, the large uncertainties of these
measurements left ambiguities.

The first DRF experiment using highly charged ions at a storage
ring was carried out with Si$^{11+}$ ions by Bartsch et
al.\cite{bar97}. It produced results with an unprecedented
accuracy, enabling a detailed comparison with theory. Whereas the
overall agreement between experiment and theory for the magnitude
of the effect (up to a factor of 3 when increasing the field from
0 V/cm to 183 V/cm) was fair, discrepancies remained in the
functional dependence of the rate enhancement on the electric
field strength. This finding stimulated theoretical investigations
of the role of the additional magnetic field which is always
present in storage ring DR experiments, since it is needed to
guide and confine the electron beam within the electron cooler. In
a model calculation Robicheaux and Pindzola\cite{rob97} found that
in a configuration of crossed $\vec{E}$ and $\vec{B}$ fields
indeed the magnetic field influences  through the mixing of $m$
levels the rate enhancement generated by the electric field. More
detailed calculations\cite{gri98,rob98} confirmed these results.
It should be noted that in theoretical calculations by Huber and
Bottcher\cite{HB80} no influence of a pure magnetic field of at
least up to 5 T on DR was found.

Inspired by these predictions we previously performed storage ring
DRF experiments using Li-like Cl$^{14+}$\cite{BSM99} and
Ti$^{19+}$\cite{bar99} ions and crossed $\vec{E}$ and $\vec{B}$
fields where we clearly discovered a distinct effect of the
magnetic field strength on the magnitude of the DR rate
enhancement. The electric field effect decreased monotonically
with the $\vec{B}$ field increasing from 30~mT to 80~mT. A
decrease of the electric field enhancement by a crossed magnetic
field is also predicted by the model calculation of Robicheaux and
Pindzola\cite{rob97} for magnetic fields larger than approximately
20 mT where, due to a dominance of the magnetic over the electric
interaction energy, the $\ell$-mixing weakens and consequently the
number of states participating in DR decreases. At lower magnetic
fields $m$-mixing yields an increase of the DR rate with
increasing $\vec{B}$ field. A corresponding experimental
observation has been made recently by Klimenko and
coworkers\cite{KKG99} who studied recombination of Ba$^+$ ions
from a continuum of finite bandwidth which they had prepared by
laser excitation of neutral Ba atoms. For a given electric field
strength of 0.5 V/cm, they find that the recombination rate is
increasingly enhanced by crossed magnetic fields up to about 20
mT. However, there is no effect of the magnetic field when it is
directed parallel to the electric field vector. For the $m$-mixing
to occur the crossed $\vec{E}$ and $\vec{B}$ arrangement is
essential. In the case of parallel $\vec{B}$ and $\vec{E}$ fields
$m$ remains a good quantum number and no influence of the magnetic
field is expected.

The aim of the present investigation with Li-like Ni$^{25+}$ is to extend
the previous studies to an ion with even higher nuclear charge $Z$.
Because of the $Z^4$ scaling of radiative rates it is expected that with
higher $Z$ less $\ell$ states of a given Rydberg $n$ level take part in DR
and therefore the sensitivity to $\ell$-mixing
decreases\cite{Mueller86/87}. Results of Griffin and Pindzola\cite{gri87}
who calculated decreasing DR rate enhancements for increasing charge
states of iron ions point into the same direction. Another aspect of going
to higher $Z$ is that even high lying $2p_j n\ell$ Rydberg resonances are
more separated in energy and therefore easier to resolve. This has been
demonstrated by Brandau et al.\ who resolved $2p_{1/2}n\ell$
DR resonances up to $n=41$ in the recombination spectrum of Li-like
Au$^{76+}$ ions\cite{BBD98}. When choosing Ni$^{25+}$ we hoped to be able
to study the field enhancement effect on a single $2p_jn\ell$ resonance.
This would enable a quantitative comparison with theory which at present
is limited to a single low value of $n$\cite{gri98} when explicitly treating
all $n\ell m$ levels required for a realistic description of DR in
crossed $\vec{E}$ and $\vec{B}$ fields.

In our DRF studies we have chosen Li-like ions as test systems since on
the one hand their electronic structure is simple enough to be treated
theoretically on a high level of sophistication and on the other hand
provides strong DR channels connected to the $2s\to2p_j$ core excitations.

\section{Experiment}\label{sec:exp}

The measurements have been performed at the heavy ion storage ring
TSR\cite{JKA89} of the Max-Planck-Institut f\"{u}r Kernphysik in
Heidelberg. For a general account of experimental techniques at
heavy ion storage rings the reader is referred to an article by
M\"{u}ller and Wolf\cite{MW97b}. Recombination measurements at storage
rings have been reviewed recently by M\"{u}ller\cite{mue99},
Schippers\cite{schi99} and Wolf et al.\cite{WGL99}. Detailed
descriptions of the experimental procedure for field free DR
measurements have been given by Kilgus et al.\cite{Kil92} and more
recently by Lampert et al.\cite{Lam96}. Therefore, we here only
describe more explicitly experimental aspects pertaining
especially to the present investigation.

The $^{58}$Ni$^{25+}$ ion beam was supplied by the MPI tandem booster facility and injected
into the TSR with an energy of 343 MeV. Using multiturn injection and e-cool
stacking\cite{Gri90} ion currents of up to 3300 $\mu$A were stored in the TSR. At these high
ion currents, however, intra beam scattering heated the ion beam during DR measurements
resulting in a considerable loss of energy resolution. To avoid this and to limit the
recombination count rate to below 1 MHz, i.~e.\ to below a count rate where dead time
effects are still negligible, we kept ion currents below 1 mA during all DR measurements. In
the storage ring the circulating Ni$^{25+}$ ions were merged with the magnetically guided
electron beam of the electron cooler. In the present experiment the electron density was
$5.4\times 10^6$~cm$^{-3}$ at cooling energy. Generally the electron density varies with the
cathode voltage $U_{\rm c}$ thereby following a $U_{\rm c}^{3/2}$ dependence. The
distribution of collision velocities in the electron-ion center of mass frame can be
described by the anisotropic Maxwellian
\begin{eqnarray}
f(\vec{v},v_{\rm rel}) &=& \frac{m_{\rm e}}{2\pi k_{\rm B}T_\perp}\,
           \exp\left(-\frac{m_{\rm e} v_\perp^2}
            {2 k_{\rm B}T_\perp}\right)\nonumber\\
           &&\times\left[\frac{m_{\rm e}}{2\pi k_{\rm B}T_{||}}\right]^{1/2}
     \exp\left(-\frac{m_{\rm e} (v_{||}-v_{\rm rel})^2}{2 k_{\rm B}T_{||}}\right)
     \label{eq:fele}
 \end{eqnarray}
characterized by the longitudinal and transverse temperatures
$T_{||}$ and $T_\perp$. In Eq.~(\ref{eq:fele}) $m_{\rm e}$ is the
electron mass, $k_{\rm B}$ is the Boltzmann constant, and $v_{\rm
rel}$ is the detuning of the average longitudinal electron
velocity from that at cooling, which determines the relative
energy $E_{\rm rel}\approx m_{\rm e}v_{\rm rel}^2/2$ between the
electron and the ion beam. The longitudinal temperature, inferred
from the experimental resolution for relative energies $E_{\rm
rel} \gg k_{\rm B}T_\perp$, was $k_{\rm B}T_{||} \approx
0.25$~meV. It implies an energy resolution given by $\Delta
E$(FWHM)$ = 4\sqrt{\ln(2) k_{\rm B}T_{||} E_{\rm
rel}}$\cite{Kil92}. The longitudinal velocity spread of the stored
ion beam yields a considerable contribution to this temperature,
while the velocity spread of the electron beam alone, after
acceleration, is estimated to be $<0.1$~meV. In the transverse
direction the electron beam was adiabatically expanded\cite{Pas96}
from a diameter $d_c \approx 9.5$~mm at the cathode to a diameter
$d_{\rm e} = 29.5$~mm in the interaction region; the reduction of
its transverse velocity spread by this expansion determines the
low value of the transverse temperature of $k_{\rm B}T_\perp
\approx 10$~meV.

Before starting a measurement, the ion beam was cooled for 5
seconds until the beam profiles reached their equilibrium widths.
This can be monitored online by employing beam profile monitors
based on residual gas ionization\cite{Hoc94}. The cooled ion beam
had a diameter $d_{\rm i} \approx 2$~mm. During the measurement
the electron cooler voltage was stepped through a preset range of
values different from the cooling voltage, thus introducing
non-zero mean relative velocities between ions and electrons.
Recombined Ni$^{24+}$ ions were counted as a function of the
cooler voltage with a CsI-scintillation detector\cite{Mie96}
located behind the first dipole magnet downstream of the electron
cooler. The dipole magnet bends the circulating Ni$^{25+}$ ion
beam onto a closed orbit and separates the recombined Ni$^{24+}$
ions from that orbit.

Two different measurement schemes were applied for the measurement
of i) a high resolution `field-free' DR spectrum (residual stray
electric fields $\leq 5$~V/cm) and ii) DRF spectra with motional
electric fields ranging up to 300 V/cm. In view of the result of
Huber and Bottcher\cite{HB80} who calculated that purely magnetic
fields below 5~T do not influence DR, the use of the term
`field-free' seems justified in case i) even with the magnetic
guiding field (up to 80~mT) still present in the electron cooler.

\subsection{Procedure for a field free high resolution measurement\label{sec:ff}}

In between two measurement steps for different values of $E_{\rm
rel}$ the cooler voltage was first set back to the cooling value
in order to maintain the ion beam quality and then set to a
reference value which is chosen to lead to a relative velocity
where the electron-ion recombination signal is very small,
favourably being only due to a negligible RR contribution
(reference relative energy $E_{\rm ref}$). Under this condition
the recombination rate measured at the reference point monitors
the background signal due to electron capture from residual gas
molecules. Choosing short time intervals of the order of only
10~ms for dwelling on the measurement, cooling and reference
voltages ensured that the experimental environment did not change
significantly  between the signal and the background measurements.
An additional interval of 1.5~ms after each change of the cooler
voltage allowed the power supplies to reach the preset values
before data taking was started.

The electron-ion recombination coefficient
 \begin{equation}
 \alpha(E_{\rm rel}) = \int\!\! d^3 \vec{v}\, \sigma(v) v f(\vec{v},v_{\rm rel})
 \label{eq:alpha}
 \end{equation}
is obtained from the background corrected recombination count rate
$R(E_{\rm rel})-R(E_{\rm ref})$, the detection efficiency $\eta$, the electron
density $n_{\rm e}$, the number of stored ions $N_{\rm i}$, the nominal
length $L=1.5$~m of the interaction zone and the ring circumference $C =
55.4$~m using the relation
\begin{equation}
\alpha(E_{\rm rel}) = \frac{R(E_{\rm rel})-R(E_{\rm
ref})}{\gamma_i^{-2}\eta\, n_{\rm e}(E_{\rm rel}) N_{\rm i} L/C} +
\alpha(E_{\rm ref})\frac{n_{\rm e}(E_{\rm ref})}{n_{\rm e}(E_{\rm
rel})}\label{eq:alphaexp}
\end{equation}
where $\gamma_{\rm i} = 1 + E_{\rm i}/(m_{\rm i}c^2)$ is the relativistic Lorentz factor for
the transformation between the c.~m.\ and the laboratory frames where the ions of mass
$m_{\rm i}$ have the kinetic energy $E_{\rm i}$. The detection efficiency of the
CsI-scintillation detector\cite{Mie96} used to detect the recombined ions is very close to
unity for count rates up to 2.5 MHz. The second term in Eq.~(\ref{eq:alphaexp}) is to be
added in case of a non-negligible electron-ion recombination rate at the reference energy.
We insert the theoretical RR rate at $E_{\rm ref} =131.5~eV$ which we have calculated to be
$\alpha(E_{\rm ref}) = 1.39\times 10^{-11}$~cm$^3$/s using a semi-classical formula for the
radiative recombination cross section\cite{BS57}
\begin{eqnarray}
\sigma_{RR}(E_{\rm rel}) &=& 2.1\times10^{-22}{\rm cm}^2 \nonumber\\
 &&\times \sum_{n_{\rm min}}^{n_{\rm cut}} k_n t_n \,\frac{q^4 {\cal R}^2}{n
 E_{\rm rel} (q^2{\cal R}+n^2 E_{\rm rel})}\label{eq:sigmarr}
\end{eqnarray}
with ${\cal R}$ denoting the Rydberg constant and $k_n$ being correction factors given by
Andersen and Bolko\cite{AB90}. This expression for recombination on bare nuclei is used to
approximately describe RR on a lithium-like core by introducing the lowest quantum number
$n_{\rm min}=2$ and weight factors $t_n$ accounting for partial occupation of $n$-shells. In
our calculation we use $t_2 = 7/8$ and $t_n = 1$ for $n > 2$. For the maximum (cut off)
quantum number we use $n_{\rm cut}=150$ as explained below.

After the generation of a recombination spectrum from the
experimental data via Eq.~(\ref{eq:alphaexp}) a correction
procedure accounting for non-perfect beam overlap in the merging
sections of the cooler is applied\cite{Lam96} which in our case
only slightly redistributes the DR resonance strengths, resulting
in DR peaks narrower and taller by small amounts. The systematic
uncertainty in the absolute recombination rate coefficient is due
to the ion and electron current determination, the corrections
accounting for the merging and demerging sections of the electron
and ion beams, and the detection efficiency. It is estimated to be
$\pm 15\%$ of the measured recombination rate
coefficient\cite{Lam96}. The statistical uncertainty of the
results presented below amounts to less than $1\%$ of the rate
coefficient maximum.

\subsection{Procedure for DRF measurements\label{sec:drf1}}

\begin{figure}
\epsfxsize=8.5cm \centerline{\epsfbox{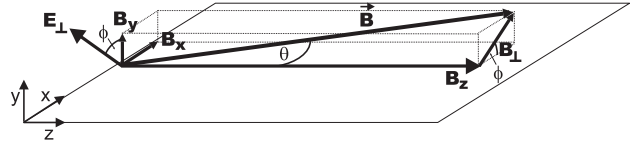}}

\caption[]{Sketch of the electric and magnetic field configuration used in DRF measurements.
The ion beam is aligned along $B_z$. The motional electric field is $E_\perp = v_iB_\perp =
v_i \sqrt{B_x^2+B_y^2}$ with the ion velocity $v_i$. The azimuthal angle $\phi$ denotes the
direction of $E_\perp$ in the $x$-$y$ plane. The electron beam is aligned along the
resulting $\vec{B}$ vector which is inclined by the angle $\theta$ with respect to $B_z$.
 \label{fig:geo} }
\end{figure}

The geometry of the magnetic and electric fields present in the
merging section of the electron cooler is sketched in
Fig.~\ref{fig:geo}. We choose the $z$-axis to be defined by the
ion beam direction. The magnetic guiding field $\vec{B}$ defines
the electron beam direction. The field strength $B$ is limited
both towards low and high values. Only fields $B > 25$ mT
guarantee a reliable operation of the electron cooler. The maximum
tolerable current through the generating coils limits $B$ to at
most 80 mT. Correction coils allow the steering of the electron
beam in the $x$-$y$ plane. In the first place these are used to
minimize the transverse field components $B_x$ and $B_y$ with
respect to the ion beam, such that the two beams are collinear and
centered to each other. The collinearity is inferred indirectly
from beam profile measurements of the cooled ion beam \cite{Hoc94}
with an accuracy of $\sim 0.2$~mrad; i.~e.\ the transverse
magnetic field components caused by imperfections in the beam
alignment amounts at most to 2$\times$10$^{-4} B$. Residual fields
which may vary in size and direction along the overlap length, are
expected to be also of this magnitude. Since the settings of the
various steering magnets result from a rather tedious beam
optimization process, they are not exactly reproduced after each
optimization procedure that is required e.~g.\ after a change of
the magnetic guiding field $B_z$. This means that the residual
transverse magnetic fields for the collinear geometry may also
slightly vary from one set of cooler settings to another. All
uncertainties in the transverse magnetic field translate into an
uncertainty in the motional electric field of less than
$\pm$10~V/cm in our present experiment.

In DRF measurements we offset the current through the correction
coils to generate additional magnetic field components $B_x$ and
$B_y$. Their influence on the stored ion beam is negligible,
i.~e.\ the ion beam is still travelling with velocity $v_{\rm i}$
in $z$-direction. However, in the frame of the ion beam the
magnetic field components $B_x$ and $B_y$ generate a motional
electric field $E_\perp = \sqrt{E_y^2+ E_x^2} = v_{\rm i}
\sqrt{B_x^2 + B_y^2}$ in the $x$-$y$ plane rotated out of the $y$
direction by the azimuthal angle $\phi = \arctan(B_y/B_x) =
\arctan(E_x/E_y)$, i.~e. $E_\perp = E_y$ for $\phi=0$. The
electrons (due to their much lower mass) follow the resulting
magnetic field vector $\vec{B}$ which now crosses the ion beam at
the angle $\theta = \arctan(\sqrt{B_x^2+B_y^2}/B_z)$. Two
consequences are to be dealt with: i) The ion beam now probes
different portions of the space charge well of the electron beam.
This reduces the energy resolution. In order to minimize this
effect we used a rather small electron density of only $5.4\times
10^6$~cm$^{-3}$ at cooling, i.~e. one order of magnitude smaller
than in the Si$^{11+}$ experiment of Bartsch et al.\cite{bar97}.
ii) The angle $\theta$ between electron beam and ion beam
explicitly enters the formula for the transformation from the
laboratory system to the c.~m.\ system which is easily derived
from the conservation of four-momentum. It reads
\begin{equation}
E_{\rm rel} = m_{\rm i} c^2 (1 + \mu) \left[\sqrt{1+\frac{2\mu}{(1+\mu)^2}
               (\Gamma-1)}-1\right]
\label{eq:cm}
\end{equation}
with the mass ratio $\mu = m_{\rm e}/m_{\rm i}$,
\begin{equation}
 \Gamma  =  \gamma_{\rm i} \gamma_{\rm e} - \sqrt{(\gamma_{\rm i}^2-1)(\gamma_{\rm e}^2-1)}
 \cos{\theta}\\
\end{equation}
and $\gamma_{\rm e} = 1 + E_{\rm e}/m_{\rm e} c^2$; $E_{\rm e}$
denotes the electron laboratory energy. It is obvious that the
cooling condition $E_{\rm rel} = 0$ can only be reached for
$\gamma_{\rm i} = \gamma_{\rm e}$ and $\theta = 0$, i.~e.\ for
$\Gamma=1$. In regular field-free measurements a scheme of
intermittent cooling is used during data taking, i.~e.\ after each
measured energy a cooling interval ($E_{\rm rel}=0$) is inserted.
For DRF measurements this would require rapid switching from
$\theta \neq 0$ to cooling with $\theta =0$. It turned out that
such a procedure heavily distorts the electron beam mainly because
of the slow response of the power supplies controlling the
steering coils. Under such conditions useful measurements could
not be performed. Therefore, we omitted the intermittent cooling
and reference measurement intervals, thereby losing resolution.
After each injection into the ring and an appropriate cooling time
the correction coils were set to produce a defined $E_\perp$ and
the cathode voltage was ramped very quickly through a preset range
with a dwell-time of only 1 ms per measurement point. In such a
manner a spectrum for one $E_\perp$ setting was collected within
only 4 s. After termination of the voltage ramp the correction
coils were set back to $\theta=0$ and the whole cycle started
again with the injection of ions into the ring. In subsequent
cycles a range of typically 30 preset $E_\perp$ values was
scanned. Each spectrum was measured as many times as needed for
reaching a satisfying level of statistical errors. This whole
procedure was repeated for different settings of the guiding field
strength $B_z$.

In order to compare only contributions from DR to the measured spectra we
subtracted an empirical background function $\alpha_{\rm BG}(E_{\rm rel})
= a_0 + a_1 E_{\rm rel} + a_2/(1+a_3 E_{\rm rel} + a_4 E_{\rm rel}^2)$
with the coefficients $a_i$ determined by fitting $\alpha_{\rm BG}(E_{\rm
rel})$ to those parts of the spectrum which do not exhibit DR resonances.
One should note that a proper calculation of the RR rate coefficient is
hampered by the fact that for $\theta\neq 0$ the electron velocity
distribution probed by the ion beam cannot be described by
Eq.~(\ref{eq:fele}).

\section{Results and Discussion}

\subsection{Recombination at zero electric field\label{sec:resdr0}}

\subsubsection{DR cross section}

For the $\Delta n = 0$ DR channels of Li-like Ni$^{25+}$, i.~e.\ for DR
involving excitations which do not change the main quantum
number of any electron in the $1s^22s$ core,
Eq.\ (\ref{eq:DR}) reads more explicitly
 \begin{eqnarray}
  e^- &+& {\rm Ni}^{25+}(1s^22s_{1/2}) \to {\rm Ni}^{24+}(1s^2 2p_jn\ell)
 \nonumber\\
 &\to&\left\{
 \begin{array}{ll}
  {\rm Ni}^{24+}(1s^2 2s_{1/2} n\ell)+ h\nu & \mbox{\rm (type I)}\\
  {\rm Ni}^{24+}(1s^2 2p_j n'\ell') + h\nu' & \mbox{\rm (type II)}
  \end{array}
  \right.
  \label{eq:DRLilike}
 \end{eqnarray}
The lowest Rydberg states which are energetically allowed are $n=13$ and
$n=11$ for $2s_{1/2}\to 2p_{1/2}$ and $2s_{1/2}\to 2p_{3/2}$ core
excitations, respectively.

\begin{figure}
\epsfxsize=8.5cm \centerline{\epsfbox{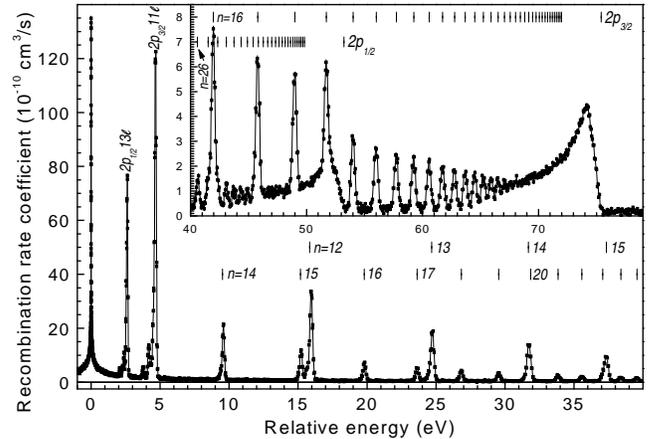}}

\caption[]{Absolute recombination rate coefficient measured for
340~MeV Ni$^{25+}$ ions. The sharp peak at zero relative energy is
due to RR. Energetic positions of the $2p_{1/2}\,nl$ and
$2p_{3/2}\,nl$ resonances according to the Rydberg formula are
indicated. \label{fig:overview} }
\end{figure}

The Ni$^{25+}(1s^22s_{1/2})$ recombination spectrum has been
measured for $0\leq \vert E_{\rm rel}\vert \leq 131.5$~eV. The
result is shown in Fig.~\ref{fig:overview}. At $E_{\rm rel} = 0$ a
sharp rise of the recombination rate due to RR is observed. At
higher energies DR resonances due to $\Delta n = 0$ $2s_{1/2}\to
2p_j$ transitions occur, the lowest resonance appearing at $E_{\rm
rel} = 0$~eV. Individual $2p_jn\ell$ resonances are resolved for
$n\leq 32$. Their resonance strengths have been extracted from the
measured spectrum by first subtracting the theoretical
recombination rate coefficient due to RR (cf.\ section
\ref{sec:ff}) where the c.~m.\ velocity spread can be neglected
since $E_{\rm rel}$ is very large compared to $kT_\perp$ and
$kT_\|$. In principle the resulting rate coefficient should be
zero at off-resonance energies. However, we find that probably due
to our approximate treatment of RR (cf.\ Eq.~(\ref{eq:sigmarr}))
small non-zero rate coefficients remain after subtraction of the
calculated RR rate coefficient. These are removed by further
subtracting a smooth background before the observed DR resonance
structures are fitted by Gaussians. (Details of the observed
smooth RR rate were not further investigated in the present work.)
The resulting values for resonance positions and strengths are
listed in Table~\ref{tab:strength}. The $2p_{1/2}13\ell$ and
$2p_{3/2}11\ell$ resonances at about 2.5~eV and 4.5~eV,
respectively, exhibit a splitting due to the interaction of the
$n\ell$-Rydberg electron with the $1s^22p_j$ core. For higher $n$
resonances this splitting decreases and cannot be observed because
of the finite experimental energy spread which increases as
$\sqrt{E_{\rm rel}}$. The $2p_j$ Rydberg series limits,
$E_\infty$, are obtained from a fit of the resonance positions
$E_n$ with $n\geq 16$ to the Rydberg formula
\begin{equation}
E_n = E_\infty - {\cal R}\left(\frac{q}{n-\delta}\right)^2 \label{eq:ryd}
\end{equation}
with the quantum defect $\delta$ as a second fit parameter. The
fit results are listed in Table~\ref{tab:slim}, where
spectroscopic values\cite{Hin89} for the series limits are also
given. Our values agree with the spectroscopic values within
$0.6\%$, i.~e.\ within the experimental uncertainty of the energy
scale\cite{Kil92}. The result that the fitted quantum defects are
almost zero reflects the fact that the interaction between the
core electrons and the Rydberg electron is weak.

The measured DR rate decreases already below the $2p_{1/2}$ and $2p_{3/2}$ series limits as
obtained from the fit to the peak positions. (cf.\ Fig.~\ref{fig:overview}). This
discrepancy results from field ionization of high Rydberg states with $n > n_{\rm f}\approx
(3.2\times 10^8 {\rm V/cm~} q^3/ E_{\rm dip})^ {1/4}$ in the charge analyzing dipole magnet
(see Sec.~\ref{sec:exp}) with magnetic field strength $B_{\rm dip} = 0.71$~T where the
moving ion experiences the motional electric field $E_{\rm dip}=v_{\rm i} B_{\rm dip}$
($n_{\rm f}$ is the classical field ionization limit). A more realistic value for the
cut-off has to account for Stark splitting and tunnelling effects. In
Ref.~\cite{Mueller86/87} an approximate value $n_{\rm cut} \approx (7.3\times 10^8 {\rm
V/cm~} q^3/ E_{\rm dip})^{1/4}$ was found. For the calculation of the actual cut-off quantum
number $n_{\rm cut}$ relevant in this experiment one has to take into account that on the
way from the cooler to the dipole magnet states with $n>n_{\rm f}$ may radiatively decay to
states below $n_f$. An approximate calculation of this effect\cite{MW97} yields $n_{\rm cut}
=150$ for the present case.

\begin{figure}
\epsfxsize=8.5cm \centerline{\epsfbox{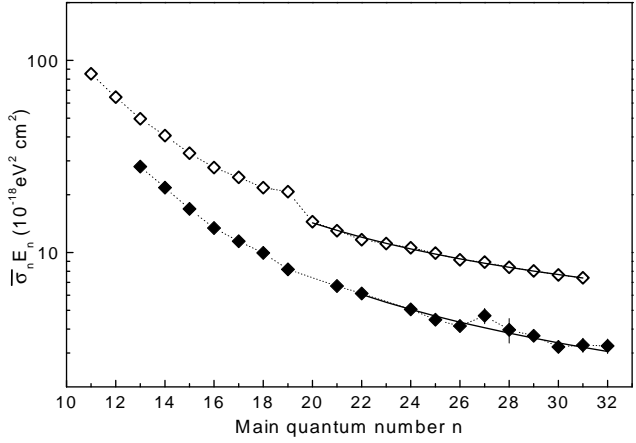}}

\caption[]{Product of strength and resonance energy of the
$2p_jn\ell$ DR resonances for $j=1/2$ (closed symbols) and $j=3/2$
(open symbols) as extracted from the experimental recombination
spectrum. Statistical error bars are mostly smaller than the
symbol size. The strengths of the $2p_{3/2}14\ell$ and
$2p_{3/2}15\ell$ resonances have been obtained by subtracting
interpolated values for the $2p_{1/2}20\ell$ and $2p_{1/2}23\ell$
resonance strengths from the measured peak areas at 31.7~eV and
37.3~eV, respectively. The thick full curves represent fits
according to Eq.~(\protect\ref{eq:sigmamodel}) with the fit
parameters listed in
Table~\protect\ref{tab:slim}.\label{fig:strength}}
\end{figure}

An estimate of the DR line strength escaping detection because of field ionization can be
made by extrapolating the measured DR line strength to $n=\infty$ employing the $n^{-3}$
scaling of the autoionization and the type II (cf.~Eq.~(\ref{eq:DRLilike})) radiative
rates\cite{WKM99}. For autoionization rates we make the ansatz $A_{\rm a}(n\ell) = A_{\rm
a}/n^3$ for $0\leq\ell\leq\ell_{\rm max}$ and $A_{\rm a}(n\ell)=0$ for $\ell_{\rm
max}<\ell<n$. Rates for the sum of type I and type II radiative transitions we represent as
$A_{\rm r}(n\ell)=A_{\rm r}^{\rm (I)} + A_{\rm r}^{\rm (II)}/n^3$. The same representations
of the relevant rates have already been used by Kilgus et al.\cite{Kil92} in a recombination
study of isoelectronic Cu$^{26+}$ ions. After summation over all $\ell$ substates
Eq.~(\ref{eq:sigma}) simplifies to
\begin{equation}
\bar{\sigma}_n E_n = S_0 \frac{A_{\rm a}\,[A_{\rm r}^{\rm
(I)}+n^{-3}A_{\rm r}^{\rm (II)}]}{A_{\rm a}+n^3A_{\rm r}^{\rm (I)}+A_{\rm
r}^{\rm (II)}} \label{eq:sigmamodel}
\end{equation}
with $S_0 = 2.475 (2j_{\rm c}+1)(\ell_{\rm max}+1)^2
\times10^{-30}{\rm cm}^2 {\rm eV}^2 {\rm s}$. For the statistical
weights in Eq.~(\ref{eq:sigma}) we have used $g_i = 2$ and
$g_d=2(2\ell+1)(2j_{\rm c}+1)$. Here, $j_c$ is the total angular
momentum quantum number of the core excited state i.~e. $j_c =1/2$
and 3/2 in the present case. The first step of the extrapolation
procedure consists of adjusting the model parameters such that
Eq.~(\ref{eq:sigmamodel}) fits the measured DR line strengths.
Here the values for $A_{\rm r}^{\rm (I)}$ have been taken from
atomic structure calculations\cite{Cow81} while $S_0$, $A_a$ and
$A_{\rm r}^{\rm (II)}$ were allowed to vary during the fit. It
turned out that the fit is not very sensitive to large variations
of the Auger rates $A_a$. In this situation we kept also the Auger
rates fixed at values which have been inferred from atomic
structure calculations and which are meant to be order of
magnitude estimates only. In Fig.~\ref{fig:strength} we have
plotted the measured and fitted resonance strengths (multiplied by
the resonance energy) as a function of the main quantum number.
The actual parameters used for drawing the fit curves are listed
in Table~\ref{tab:slim}. The fit has been restricted to $n \geq
20$ for both the $2p_{1/2}$ and the $2p_{3/2}$ series because
additional Coster-Kronig decay channels $2p_{3/2}n\ell \to
2p_{1/2}\epsilon\ell'$ open up when $E_n$ crosses the $2p_{1/2}$
series limit. A corresponding discontinuous decrease of the
$2p_{3/2}n\ell$ DR resonance strength can be clearly discerned in
Fig.~\ref{fig:strength}. Since the additional autoionizing
channels are included for the $2p_{3/2}n\ell$ series of resonances
but not for the $2p_{1/2}n\ell$ series, $A_{\rm a}$ for the
$2p_{3/2}n\ell$ series is more than a factor of 2 higher than
$A_{\rm a}$ for the $2p_{1/2}n\ell$ series. A factor of 2 would
just correspond to the ratio of statistical weights.

\begin{figure}
\epsfxsize=8.5cm \centerline{\epsfbox{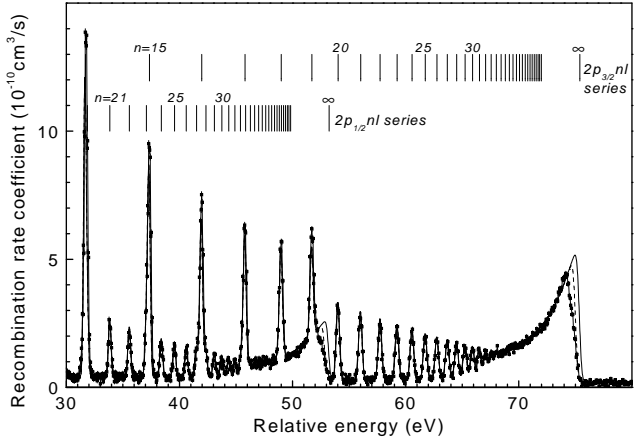}}

\caption[]{Comparison between measured (closed symbols) and extrapolated recombination
spectra (see text). The extrapolations extend to $n=150$ (dashed line) and $n=500$ (full
line) where convergence is achieved. \label{fig:simul}}
\end{figure}

Inserting the fit parameters listed in Table~\ref{tab:slim} into
Eqs.~(\ref{eq:ryd}) and (\ref{eq:sigmamodel}) now allows an
extrapolation of the DR resonance positions and strengths,
respectively, to be obtained for arbitrary high $n$. In order to
check the quality of the extrapolation we have convoluted the
extrapolated DR cross section with the experimental electron
energy distribution using the electron beam temperatures $k_{\rm
B}T_{||}=0.25$~meV and $k_{\rm B}T_\perp=10$~meV. After adding the
semiclassically calculated RR rate coefficient --- as described
above
--- the resulting extrapolated DR+RR recombination rate
coefficient is plotted in Fig.~\ref{fig:simul} together with the experimental one. Despite
the very simple model assumptions the calculated recombination rate agrees with the measured
one also over the energy intervals covered by the $2p_jn\ell$ resonances with $n>31$ which
are not resolved individually and therefore have not been used for the fits. Deviations for
the Ni$^{24+}(1s^22p_{3/2}n\ell)$ resonances with $n\leq 19$ stem from the fact that for
these resonances the Coster-Kronig decay channels to Ni$^{25+}(1s^2 2p_{1/2})$ are closed
whereas the fit has been made to resonances where they are open. At energies close to the
series limits slight deviations from the model rate occurs even when the expected value of
$n_{\rm cut}=150$ is inserted into the model (dashed line in Fig.~\ref{fig:simul}). The
origin of this discrepancy has probably to be searched for in the approximations made in
both the field ionization model and in the model rate descriptions, in particluar regarding
the dependence of the angular momentum $\ell$ which may be affected by even the small
residual electric fields in the interaction region.

\subsubsection{Maxwellian plasma rate coefficient}

The comparison between the measured data and the calculated extrapolation to $n\to\infty$
(full line in Fig.~\ref{fig:simul}) suggests that only a minor part of the total Ni$^{25+}$
$\Delta n=0$ DR resonance strength has not been measured. This enables us to derive from our
measurement the Ni$^{25+}$ DR rate coefficient in a plasma. To this end the experimental DR
rate coefficient is substituted by the extrapolated one at energies $\sim 3$~eV below the
series limits. Compared to using the experimental result without extrapolation this results
in a correction of the plasma rate coefficient of at most $5\%$. The experimental DR rate
coefficient including the high $n$-extrapolation is convoluted by an isotropic Maxwellian
electron velocity distribution characterized by the electron temperature $T_{\rm e}$. The
resulting $\Delta n=0$ DR plasma rate coefficient is displayed in Fig.~\ref{fig:plasma}
(thick full line). Summing experimental and extrapolation errors, the total uncertainty of
the DR rate coefficient in plasmas determined in this work amounts to $\pm 20\%$.

A convenient representation of the plasma DR rate coefficient is provided
by the following fit formula
\begin{equation}
\alpha(T_{\rm e}) = T_{\rm e}^{-3/2} \sum_i c_i\exp{(-E_i/k_{\rm B}T_{\rm
e})} \label{eq:alphafit}
\end{equation}
It has the same functional dependence on the plasma electron
temperature as the widely used Burgess formula\cite{bur65}, where
the coefficients $c_i$ and $E_i$ are related to oscillator
strengths and excitation energies, respectively. The results for
the fit to the experimental Ni$^{25+}$ $\Delta n=0$ DR rate
coefficient in a plasma (thick full line in Fig.~\ref{fig:plasma})
are summarized in Table~\ref{tab:tfit}. The fitted curve cannot be
distinguished from the experimental plasma rate coefficient in a
plot as presented in Fig.~\ref{fig:plasma}.

In Fig.~\ref{fig:plasma} we also compare our results with
theoretical results for $\Delta n=0$ DR by Mewe et al.\cite{MSS80}
(dashed line), Romanik\cite{rom88} (dash-dotted line) and Teng et
al.\cite{TSZ94} (dashed-dot-dotted line) who interpolated DR
calculations performed by Chen\cite{Che91} for selected
lithiumlike ions. At temperatures $k_{\rm B}T_{\rm e} > 1$~eV the
rate of Mewe et al., which is based on the Burgess
formula\cite{bur65}, overestimates our experimental result by up
to a factor of $\sim 5$. At lower temperatures the experimental
result is underestimated by factors up to 10. Above an electron
temperature of 30 eV Romanik's theoretical result agrees with our
$\Delta n=0$ DR rate coefficient to within $15\%$, which is within
the $20\%$ experimental accuracy. In this energy range the
interpolation result of Teng et al. underestimates the
experimental rate coefficient by 20--30$\%$. It should be noted
that neither the calculation of Romanik nor that of Teng et al.\
covers temperatures below 10~eV. When compared with our RR
calculation (thin full line in Fig.~\ref{fig:plasma}) our
experimental result shows that in the temperature range of 1 to
10~eV, where Ni$^{25+}$ ions may exist in photoionized plasmas, DR
is still significant. The importance of DR in low temperature
plasmas has been pointed out recently by Savin et al.\cite{SBC97}
who measured DR of fluorine-like Fe$^{17+}$ ions. At higher
temperatures Romanik's calculation\cite{rom88} suggests that above
100~eV $\Delta n=1$ DR contributions become significant (upper
dash-dotted line in Fig.~\ref{fig:plasma}).

\subsection{DRF measurements}

\begin{figure}
\epsfxsize=8.5cm \centerline{\epsfbox{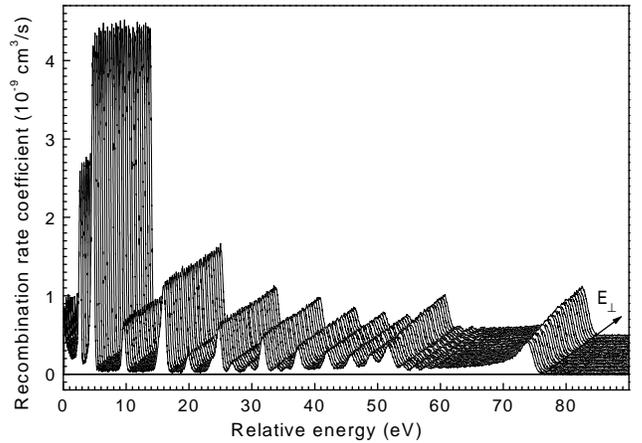}}

\caption[]{Ni$^{25+}$ DR spectra for 28 external electric fields 0~$\leq
E_\perp\leq$~270~V/cm and $\vert \cos\phi\vert=1$. Spectra with $\phi=0^\circ$ and
$\phi=180^\circ$ are interleaved. Adjacent spectra differ by $\Delta E_\perp = 10$~V/cm. The
magnetic field on the axis of the electron cooler has been $B_z=80$~mT. The fitted smooth RR
contribution has been subtracted.\label{fig:fields} }
\end{figure}

\begin{figure}
\epsfxsize=8.5cm \centerline{\epsfbox{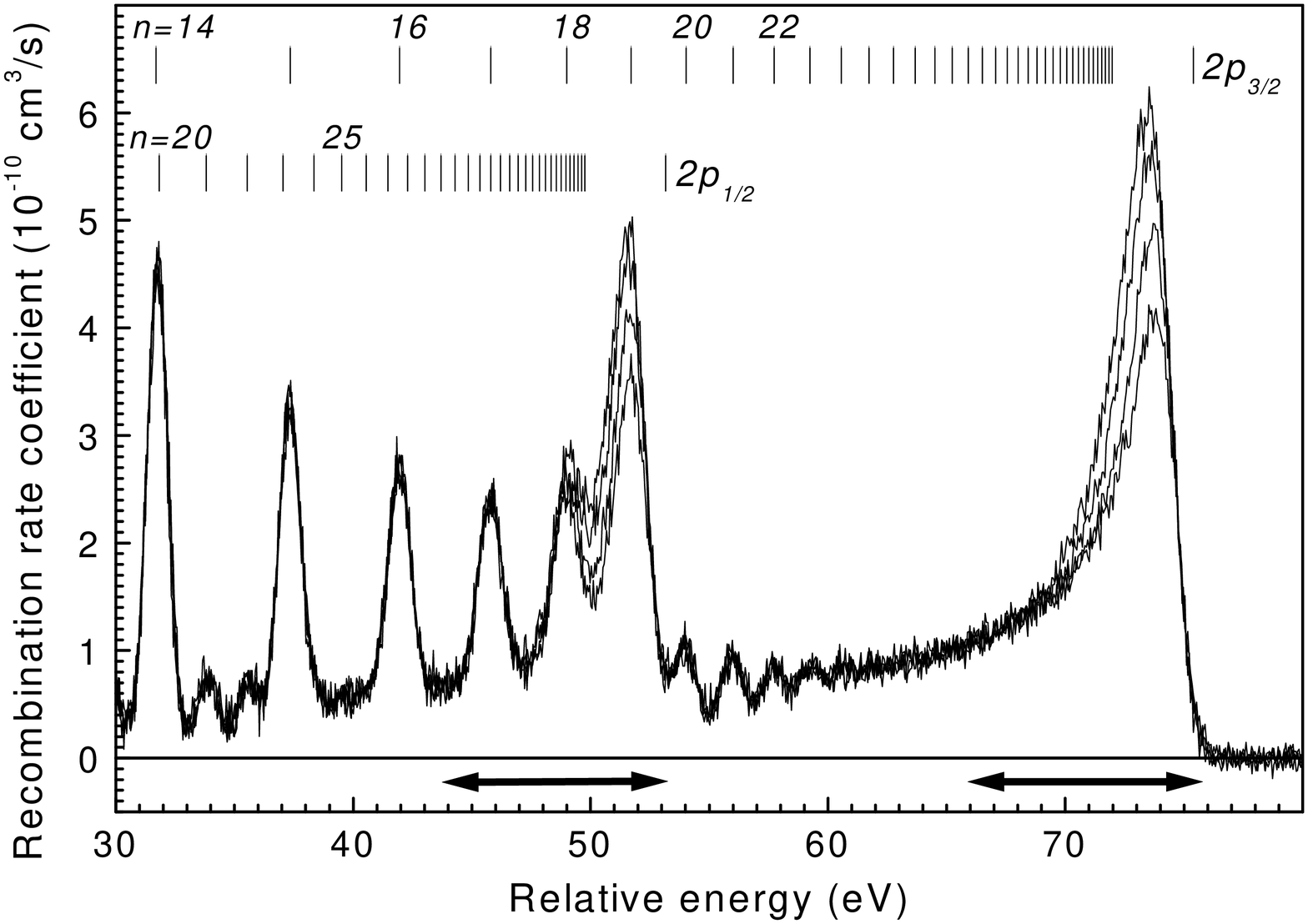}}%

\caption[]{Ni$^{25+}$ DR spectra for 4 external electric fields 0, 110, 190 and 270~V/cm
clearly showing the DR rate enhancement with increasing electric field strength close to the
$2p_{1/2}$ and $2p_{3/2}$ series limits. Ranges for the determination of integrated rate
coefficients are indicated by horizontal arrows. The magnetic field on the axis of the
electron cooler has been $B_z=80$~mT. The fitted smooth RR contribution has been
subtracted.\label{fig:grow} }
\end{figure}

Fig.~\ref{fig:fields} shows a series of Ni$^{25+}$ recombination
spectra measured in the presence of external electric fields
$E_\perp$ ranging from 0 to 270 V/cm. The magnetic field on the
axis of the cooler has been $B_z = 80$~mT. Due to the altered
measurement scheme that leaves out the intermittent cooling of the
ion beam, the energy resolution is reduced compared to
Fig.~\ref{fig:overview}. Now individual $2p_jn\ell$ DR resonances
are resolved only up to $n=21$. Two features in the series of
spectra are to be noted. Firstly, the strength of the DR
resonances occurring below 47~eV does not depend on $E_\perp$.
Secondly, the strength of the unresolved high-$n$ DR resonances
increases with increasing field strength. This can been seen more
clearly from the close-up presented in Fig.~\ref{fig:grow}. At
energies of more than 10~eV below the $2p_j$ series limits the
different DR spectra lie perfectly on top of each other whereas at
higher energies (i.~e.\ $n>30$) an increase of the DR intensities
by up to a factor of 1.5 at $E_\perp=270$~V/cm is observed. The
degraded resolution of the DR spectrum does not allow us to
resolve $n$ levels in the range of the electric field enhancement.
In order to quantify this DR rate enhancement we consider
integrated recombination rates with the integration intervals
chosen as marked in Fig.~\ref{fig:grow}.

\begin{figure}
\epsfxsize=8.5cm \centerline{\epsfbox{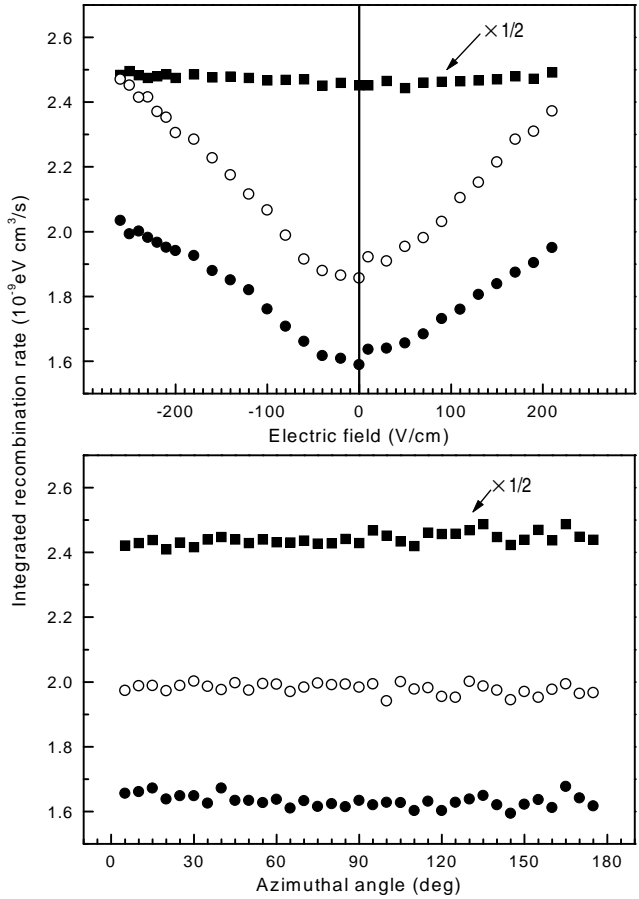}}

\caption[]{Integrated recombination rate coefficients $I_0$ (full squares), $I_{1/2}$ (full
circles) and $I_{3/2}$ (open circles) for the integration ranges defined in the text as a
function of electric field (upper panel, $\phi=0^\circ$) and of azimuthal angle (lower
panel, $E_\perp = 100$~V/cm). The magnetic guiding field is 80~mT in both
cases.\label{fig:integrals}}
\end{figure}

The integration intervals 44.0--53.5~eV and 66.2--76.0~eV include
all $2p_jn\ell$ resonances with $n\geq 31$ for $j=1/2$ and
$j=3/2$, respectively. In the following we denote the resulting
integrals by $I_{1/2}$ and $I_{3/2}$. The energy range
44.0--53.5~eV also contains $2p_{3/2}n\ell$ resonances with
$16\leq n\leq 19$. These resonances, however, are not affected by
the electric field strengths used in our experiment. Consequently,
any change in the magnitude of $I_{1/2}$ as a function of
$E_\perp$ we attribute to field effects on $2p_{1/2}n\ell$
resonances. As a check of the proper normalization of the DR
spectra we additionally monitor the integral $I_0 = \int_{2{\rm
eV}}^{18{\rm eV}}\alpha_{\rm DR}(E_{\rm rel})dE_{\rm rel}$ which
comprises the strengths of the $2p_{3/2}11\ell$, $2p_{3/2}12\ell$
and $2p_{1/2}n\ell$ DR resonances with $13\leq n \leq 15$. Since
these low $n$ resonances are not affected by $E_\perp$ we expect
$I_0$ to be constant. Any deviation of $I_0$ from a constant value
would indicate a reduction of beam overlap due to too large a
tilting angle $\theta$ of the electron beam. The maximum angle
$\theta_{\rm max}$  to which the overlap of the electron beam with
the ion beam is ensured over the full interaction length $L$ is
given by $\tan\theta_{\rm max} = (d_{\rm e}-d_{\rm i})/L$. With
the geometrical values given above one obtains $\theta_{\rm max}
\approx 1^\circ$. Apart from the highest $E_\perp$ at
$\phi=180^\circ$ and at the lowest magnetic guiding field, i.~e.\
at $B_z=41.8$~mT, the condition $\theta < \theta_{\rm max}$ was
always met. This is exemplified in the upper panel of Fig.
\ref{fig:integrals} where $I_0$, $I_{1/2}$ and $I_{3/2}$ are shown
for -270~V/cm~$\leq E_\perp \leq$~220~V/cm. There, positive
(negative) field strength indicates $\phi=0^\circ$
($\phi=180^\circ$). While $I_{1/2}$ and $I_{3/2}$ clearly exhibit
a field effect which is nearly symmetric about $E_\perp=0$~V/cm,
$I_0$ is independent of $E_\perp$.

The fact that the increase of the integrated recombination rate
coefficient is independent of the sign of the electric field
vector is expected from the cylindrical symmetry of the merged
beams arrangement in the electron cooler. However, for the entire
experimental setup this symmetry is broken by the charge analyzing
dipole magnet with an electric field vector lying in the bending
plane (the $x$-$y$ plane of the coordinate frame defined in the
interaction region). This, in principle, could lead to
redistribution of population between different $m$ substates in
the dipole magnet\cite{NH87} and a resulting field ionization
probability depending on the azimuthal angle $\phi$ of the
motional electric field vector in the cooler. In order to clarify
this question we took a series of DRF spectra with $\phi$ ranging
from 5$^\circ$ to 175$^\circ$. At the same time the electric and
magnetic fields were kept fixed at $E_\perp = 100$~V/cm and
$B_z=80$~mT. A scan around a full circle was prohibited by the
limited output of the power supplies used for steering the
electron beam in the particular arrangement of this experiment. As
shown in the lower panel of Fig.~\ref{fig:integrals} no
significant dependence of the integrated recombination rates on
the azimuthal angle $\phi$ was found.

\begin{figure}
\epsfxsize=8.5cm \centerline{\epsfbox{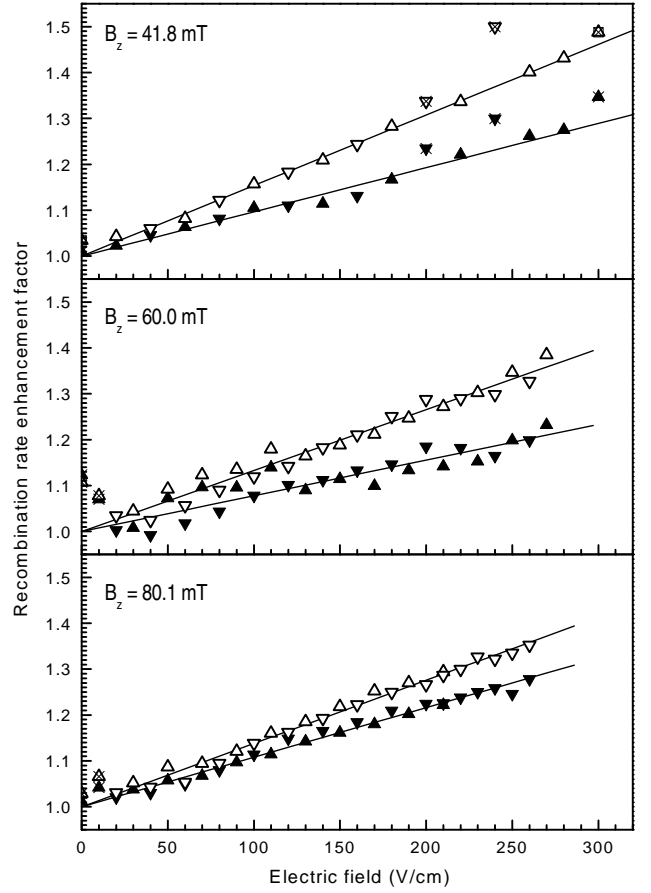}}

\caption[]{Recombination rate enhancement factors as a function of electric field $E_\perp$
for three different magnetic fields $B_z$. The enhancement is larger for the series of
$2p_{3/2}n\ell$ resonances (open symbols) as compared to the $2p_{1/2}n\ell$ series (closed
symbols). Triangles pointing upwards (downwards) mark data points measured at $\phi=0^\circ$
($\phi=180^\circ$). The data points which for experimental reasons (see text) have been
excluded from the linear fits (full curves) are marked additionally by crosses. Included and
excluded data points for $B_z = 41.8$~mT and $E_\perp \geq 200$~V/cm correspond to
$\phi=0^\circ$ and $\phi=180^\circ$, respectively.
 \label{fig:efac}}
\end{figure}

As a measure for the magnitude of the field enhancement we introduce the
field enhancement factor
 \begin{equation}
 r_j(E_\perp,B_z) = C_j(B_z)
 \frac{I_j(E_\perp,B_z)}{I_0(E_\perp,B_z)}
 \label{eq:efac}
 \end{equation}
for $j=1/2$ or $3/2$ and the constant $C_j(B_z)$ chosen such that
$r_j(0,B_z) =1.0$ (see below). Plots of enhancement factors as a
function of $E_\perp$ are shown in Fig.~\ref{fig:efac} for
different values of $B_z$. Since the field effect is independent
of the orientation of the electric field in the $x$-$y$ plane data
points for $\phi=0^\circ$ and $\phi=180^\circ$ are plotted
together. The enhancement factor exhibits a linear dependence on
the electric field. Exceptions occur for $B_z=41.8$~mT and
$\phi=180^\circ$ at $E_\perp$ values where $\theta$ becomes
maximal, and around $E_\perp=0$~V/cm. In the former case the
complete overlap of the ion beam and the electron beam over the
full length of the interaction region is lost. This is indicated
by a reduction of $I_0$, and apparently a consistent normalization
cannot be carried out. In the latter case residual electric and
magnetic field components resulting e.~g.\ from a non-perfect
alignment of the beams prevent us from reaching $E_\perp=0$~eV.
After excluding all data points with $E_\perp \leq 10$~V/cm and
those with $\phi=180^\circ$, $E_\perp \geq 200$~V/cm for
$B_z=41.8$~mT, we were able to fit straight lines to the measured
field enhancement factors as a function of $ E_\perp $. The
constants $C_j(B_z)$ in Eq.~(\ref{eq:efac}) have been chosen such
that the fitted straight lines yield $r^{\rm (fit)}_j(0,B_z)=1.0$.

\begin{figure}
\epsfxsize=8.5cm \centerline{\epsfbox{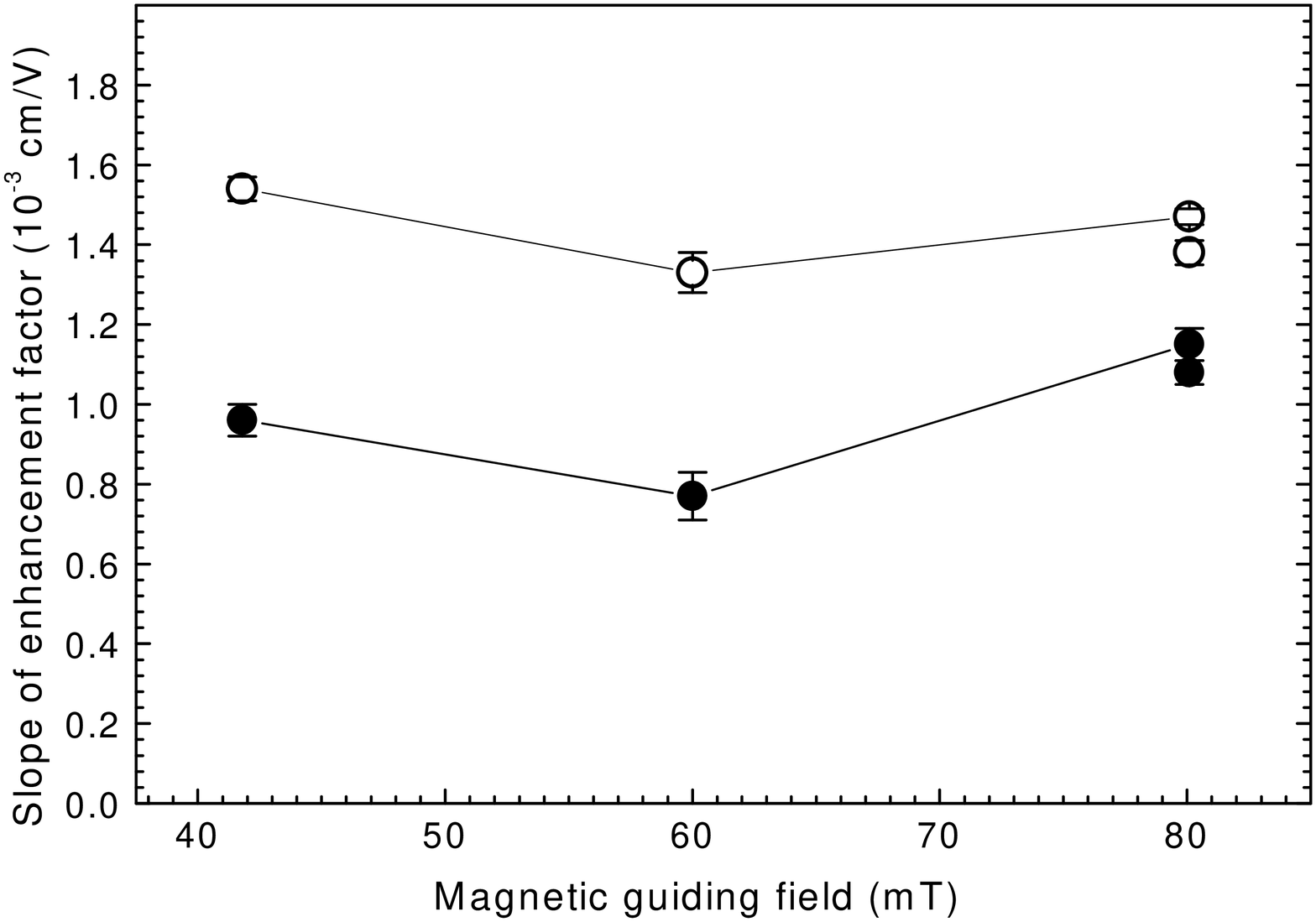}}

\caption[]{Slopes of the enhancement factors as a function of the magnetic field $B_z$ for
the $2p_{3/2}n\ell$ (open symbols) and $2p_{1/2}n\ell$ (closed symbols) series of Rydberg
resonances with $n\geq 31$.
 \label{fig:slope}}
\end{figure}

As a measure for the electric-field enhancement of the DR rate enhancement we now consider
the slopes
 \begin{equation}
 s_j(B_z) = \frac{dr^{\rm (fit)}_j(E_\perp,B_z)}{dE_\perp}
 \label{eq:slope}
 \end{equation}
of the fitted straight lines, which are displayed as a function of the magnetic field
strength in Fig.~\ref{fig:slope}. The error bars correspond to statistical errors only.
Systematic uncertainties e.~g.\ due to residual fields are difficult to estimate.
Nevertheless, their order of magnitude can be judged from the $\approx 10\%$ difference
between the two data points at $B_z =80$~mT which have been measured with different cooler
settings.

For all magnetic fields used the electric field dependence of the
rate enhancement factor is steeper for the $2p_{3/2}n\ell$ series
of Rydberg resonances than that for the $2p_{1/2}n\ell$ series.
This can be understood from the fact that the multiplicity of
states and consequently the number of states which can be mixed is
two times higher for $j=3/2$ than for $j=1/2$. Following this
argument one would expect a ratio of 2 for the respective
incremental integrated recombination rates
$I_j(E_\perp,B_z)-I_j(0,B_z)$. From our measurements we find lower
values scattering around 1.5 indicating a somewhat reduced number
of states available for field mixing within the series of
Ni$^{24+}(1s^22p_{3/2}n\ell)$ DR resonances. In calculations for
Li-like Si$^{11+}$ and C$^{3+}$ ions ratios, even less than 1 have
been found\cite{gri98,gri98b}. This has been attributed to the
electrostatic quadrupole-quadrupole interaction between the $2p$
and the $n\ell$ Rydberg electron in the intermediate doubly
excited state, which more effectively lifts the degeneracy between
the $2p_{3/2}n\ell$ than between the $2_{1/2}n\ell$ levels.
Another reason for the reduced number of $2p_{3/2}n\ell$ states
participating in DRF might be the existence of additional
Coster-Kronig decay channels for these resonances with $n\geq 20$
to Ni$^{25+}(1s^22p_{1/2})$ (cf.\ section \ref{sec:resdr0}).

\begin{figure}
\epsfxsize=8.5cm \centerline{\epsfbox{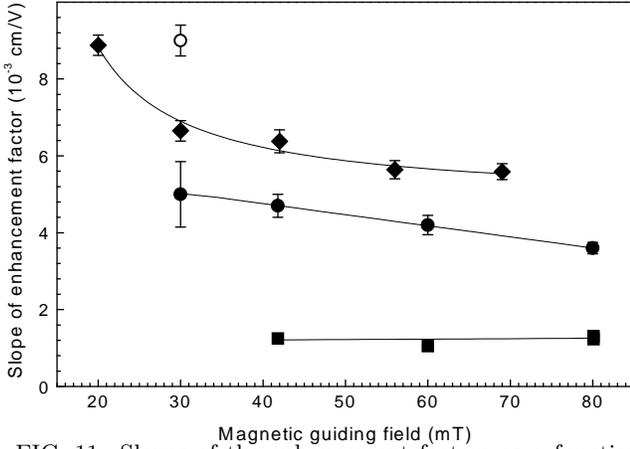}}

\caption[]{Slopes of the enhancement factors as a function of the magnetic field $B_z$ for
the $2p_{j}n\ell$ resonances of Li-like Si$^{11+}$ (open circle) \protect\cite{bar97},
Cl$^{14+}$ (diamonds) \protect\cite{BSM99}, Ti$^{19+}$ (full circles) \protect\cite{bar99}
and Ni$^{25+}$ ions (squares) [this work]. The lines are drawn to guide the eye. For
Ti$^{19+}$ and Ni$^{25+}$ where the field effects on the $2p_{1/2}$ and $2p_{3/2}$ series of
DR resonances have been determined separately, the corresponding slopes have been averaged.
Clearly the effects of external electric and magnetic fields on DR decrease with increasing
nuclear charge. \label{fig:allslopes}}
\end{figure}

Slopes for the relative electric field enhancement according to
Eq.~(\ref{eq:slope}) were previously determined in measurements by
Bartsch and coworkers for the lighter isoelectronic ions
Si$^{11+}$, Cl$^{14+}$ and Ti$^{19+}$ \cite{bar97,BSM99,bar99}.
The present results are compared with these previous data in
Fig.~\ref{fig:allslopes}. It should be noted that the comparison
is only semi-quantitative, because the choice of integration
ranges for the calculation of the integrated recombination
coefficients is somewhat arbitrary and different cut-off quantum
numbers $n_{\rm cut}$ exist for the different ions (cf.\
Table~\ref{tab:allions}). Clearly, the relative enhancement for a
given electric field strength is much lower for Ni$^{25+}$ than
for the lighter ions studied so far, with the reduction for the
step from lithium-like titanium ($Z=22$) to nickel ($Z=28$)
apparently much larger than the step from, e.~g., chlorine
($Z=17$) to $Z=22$. As a general trend with increasing $Z$, the
radiative decay rates $A_r$ decrease $\propto Z$ for type I
transitions (see Eq.~(\ref{eq:DRLilike})) and $\propto (Z-3)^4$
for type II transitions, while the autoionization rates are rather
independent of $Z$. This shifts the range where the low-$\ell$
autoionization rates are larger than the radiative stabilization
rates (the condition for DR enhancement by $\ell$-mixing to occur
\cite{rob97}) down to states of lower principal quantum number $n$
for increasing $Z$. For Ni$^{25+}$ the degree of mixing reached by
the typical experimental field strengths appears to be much
reduced as compared to lighter ions. In addition, a clear
dependence of the electric field enhancement on the the magnetic
field strength is no longer observed for Ni$^{25+}$. To clarify
the reason for the strong reduction of both the electric and the
additional magnetic field effect in the heavy system studied here,
detailed quantitative calculations are desirable.

\begin{figure}
\epsfxsize=8.5cm \centerline{\epsfbox{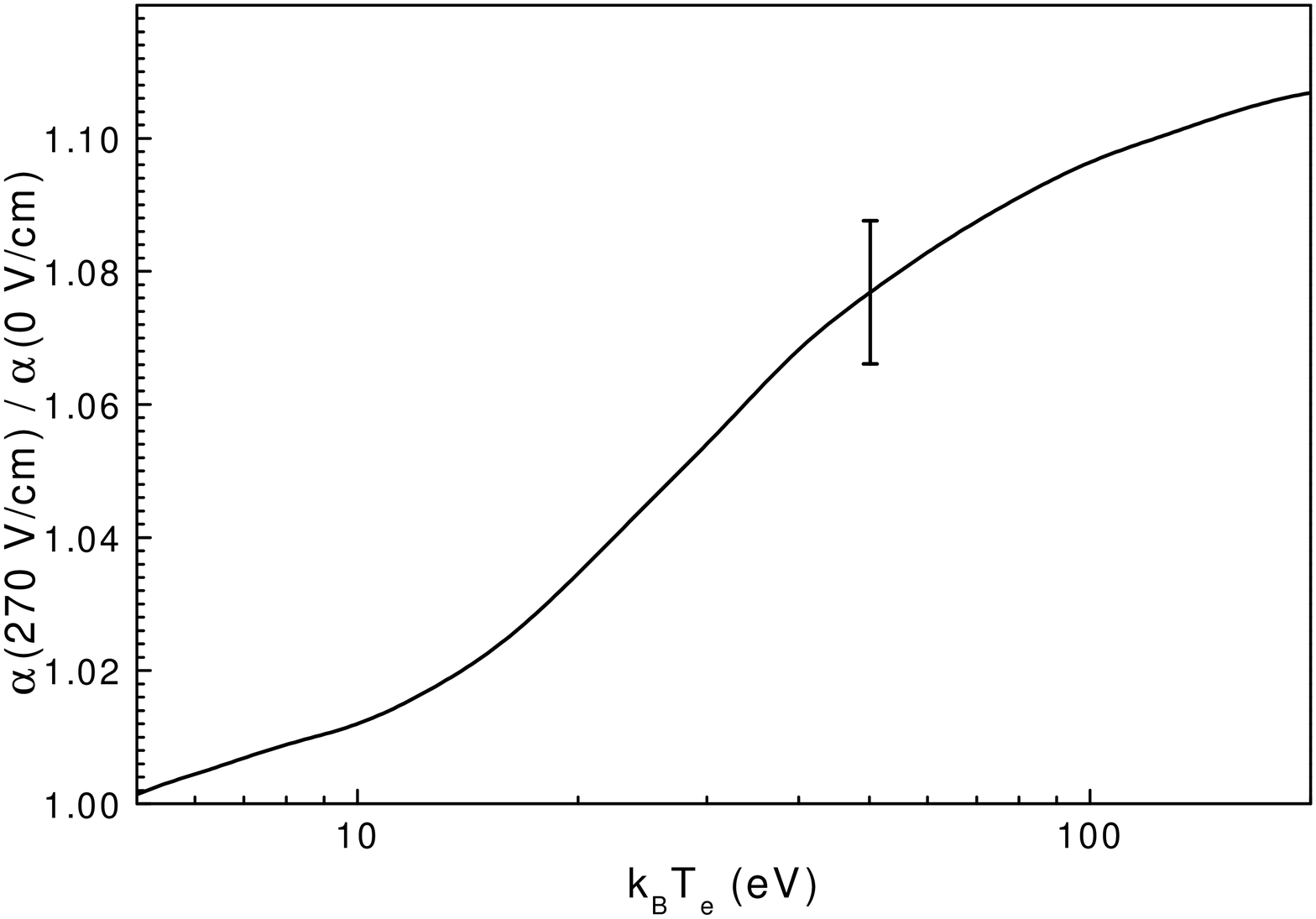}}

\caption[]{Ratio of Ni$^{25+}$ $\Delta n=0$ DR plasma rate coefficients with ($E_\perp =
270$~V/cm) and without ($E_\perp \approx 0$~V/cm ) external electric field. The magnetic
field was $B_z = 80$~mT. An experimental error bar of $\pm 1\%$ as indicated is estimated
for the ratio of the rate coefficient.\label{fig:ratio}}
\end{figure}

External electric and magnetic fields are ubiquitous in
astrophysical or fusion plasmas. Therefore, it is of interest to
look into the implications of the field enhancement effect for the
Ni$^{25+}$ $\Delta n=0$ DR rate coefficient in a plasma. As an
example we show in Fig.~\ref{fig:ratio} the ratio of rate
coefficients derived from two measurements with and without
external electric field. As compared to zero electric field, the
recombination coefficient at our highest experimental electric
field strength ($E_\perp = 270$~V/cm at $B_z=80$~mT) is enhanced
by up to 11$\%$. This value only represents a lower limit for the
enhancement at the given field strength, as we observe DR
resonances due to Ni$^ {24+}(1s^22p_jn\ell)$ intermediate states
only up to $n_{\rm cut} \approx 150$. It should also be noted that
in our experiments we did not reach the electric field strength
where the DR rate enhancement saturates.

\section{Summary and conclusions}

The recombination of lithium-like Ni$^{25+}$ ions has been
experimentally studied in detail. Spectroscopic information on
individual DR resonances associated with Ni$^{24+}(1s^22p_jn\ell)$
intermediate states, which have been experimentally resolved up to
$n=32$, has been extracted and the $\Delta n=0$ DR plasma rate
coefficient has been derived. Our experimental result is
underestimated by up to a factor of 2 by semi-empirically
calculated rate coefficients. At plasma temperatures above 10 eV,
results of detailed theoretical calculations are
available\cite{rom88} which agree well with the experiment. At
lower temperatures, where Ni$^{25+}$ ions may exist in
photoionized plasmas, no data for the DR rate coefficient have
been previously available.

In the presence of external electric fields up to 300~V/cm, the
measured DR resonance strength is enhanced by a factor 1.5; a
rather weak effect in comparison with previous measurements at $Z
= 11$, 17 and 22. Due to the overall weakness of the field effect
for Ni$^{25+}$ ions, a marked dependence of the DR rate
enhancement on the strength of a crossed magnetic field as
observed for Cl$^{14+}$ and Ti$^{19+}$ ions has not been
detectable in the present investigation.

Experimental limitations prevented us from obtaining an energy
resolution in our DRF measurements comparable to that achieved in
the field-free measurement. Consequently, we could not resolve a
Ni$^{24+}(1s^22p_jn\ell)$ DR resonance with $n$ high enough to
exhibit a field effect. Such an observation would have facilitated
a direct comparison with {\em ab initio} calculations, which due
to the large number of $n\ell m$ states to be considered,
presently can only treat a single $n$ manifold of Rydberg states
in the presence of crossed $\vec{E}$ and $\vec{B}$
fields\cite{gri98}. Improvements of this situation can be expected
in the near future from the steady increase of computing power on
the theoretical side, and on the experimental side from a
dedicated electron target which is presently being installed at
the TSR. With the electron target and the electron cooler
operating at the same time we will be able to perform DRF
measurements with continuously cooled ion beams, yielding DRF
spectra with increased resolution.

\acknowledgements

We gratefully acknowledge support by the German Federal Ministry
for Education and Research (BMBF) through contracts no.\ 06 GI 848
and no.\ 06 HD 854. R.\ P.\ acknowledges support by the Division
of Chemical Sciences, U.S.\ Department of Energy under contract
DE-FG03-97ER14787 with the University of Nevada, Reno.

\newpage
\narrowtext

\begin{table}
\begin{tabular}{d@{$\;\pm$}dd@{$\;\pm$}dl}
 \multicolumn{2}{c}{resonance position} & \multicolumn{2}{c}{resonance strength} & designation\\
 \multicolumn{2}{c}{$E_{\rm d}$ (eV)} & \multicolumn{2}{c}{$\bar{\sigma}_{\rm d}$ ($10^{-19}$eV$\,$cm$^2$)} & \\
 \hline
 2.079 & 0.002 & 2.42  & 0.12  & $ 2p_{1/2}13s $ \\
 2.267 & 0.003 & 3.25  & 0.22  & $ 2p_{1/2}13p $ \\
 2.383 & 0.001 & 9.53  & 0.26  & $ 2p_{1/2}13d $ \\
 2.625 & 0.003 & 94.2  & 5.7   & $ 2p_{1/2}13\ell (\ell>2) $ \\
 9.590 & 0.003 & 22.7  & 0.71  & $ 2p_{1/2}14\ell $\\
 15.21 & 0.001 & 11.1  & 0.12  & $ 2p_{1/2}15\ell $\\
 19.81 & 0.002 & 6.78  & 0.083 & $ 2p_{1/2}16\ell $\\
 23.63 & 0.002 & 4.85  & 0.077 & $ 2p_{1/2}17\ell $\\
 26.83 & 0.003 & 3.71  & 0.072 & $ 2p_{1/2}18\ell $\\
 29.54 & 0.003 & 2.77  & 0.057 & $ 2p_{1/2}19\ell $\\
 33.84 & 0.004 & 1.98  & 0.046 & $ 2p_{1/2}21\ell $\\
 35.57 & 0.004 & 1.72  & 0.045 & $ 2p_{1/2}22\ell $\\
 38.40 & 0.005 & 1.32  & 0.044 & $ 2p_{1/2}24\ell $\\
 39.55 & 0.009 & 1.13  & 0.065 & $ 2p_{1/2}25\ell $\\
 40.63 & 0.006 & 1.02  & 0.040 & $ 2p_{1/2}26\ell $\\
 41.53 & 0.024 & 1.13  & 0.10  & $ 2p_{1/2}27\ell $\\
 42.35 & 0.023 & 0.935 & 0.14  & $ 2p_{1/2}28\ell $\\
 43.09 & 0.008 & 0.857 & 0.042 & $ 2p_{1/2}29\ell $\\
 43.75 & 0.012 & 0.739 & 0.051 & $ 2p_{1/2}30\ell $\\
 44.35 & 0.014 & 0.744 & 0.063 & $ 2p_{1/2}31\ell $\\
 44.91 & 0.013 & 0.727 & 0.063 & $ 2p_{1/2}32\ell $\\
\hline
 3.777 & 0.002 & 4.71 & 0.13 & $ 2p_{3/2}11s $ \\
 4.190 & 0.001 & 17.9  & 0.17  & $ 2p_{3/2}11p $ \\
 4.673 & 0.001 & 163.3 & 2.1   & $ 2p_{3/2}11\ell (\ell>1) $\\
 15.96 & 0.001 & 40.4  & 0.61  & $ 2p_{3/2}12\ell $\\
 24.73 & 0.002 & 20.04 & 0.66  & $ 2p_{3/2}13\ell $\\
 31.74 & 0.001 & 15.2  & 0.14  & $ 2p_{1/2}20\ell, 2p_{3/2}14\ell$\\
 37.34 & 0.004 & 10.33 & 0.30  & $ 2p_{1/2}23\ell, 2p_{3/2}15\ell$\\
 41.95 & 0.003 & 6.61  & 0.20  & $ 2p_{3/2}16\ell $\\
 45.77 & 0.002 & 5.38  & 0.047 & $ 2p_{3/2}17\ell $\\
 48.97 & 0.002 & 4.45  & 0.045 & $ 2p_{3/2}18\ell $\\
 51.70 & 0.003 & 4.01  & 0.17 & $ 2p_{3/2}19\ell $\\
 54.00 & 0.003 & 2.68  & 0.039 & $ 2p_{3/2}20\ell $\\
 55.99 & 0.003 & 2.32  & 0.038 & $ 2p_{3/2}21\ell $\\
 57.73 & 0.004 & 2.02  & 0.038 & $ 2p_{3/2}22\ell $\\
 59.22 & 0.004 & 1.88  & 0.037 & $ 2p_{3/2}23\ell $\\
 60.56 & 0.004 & 1.75  & 0.035 & $ 2p_{3/2}24\ell $\\
 61.72 & 0.005 & 1.61  & 0.036 & $ 2p_{3/2}25\ell $\\
 62.75 & 0.005 & 1.46  & 0.034 & $ 2p_{3/2}26\ell $\\
 63.68 & 0.005 & 1.40  & 0.035 & $ 2p_{3/2}27\ell $\\
 64.49 & 0.005 & 1.30  & 0.037 & $ 2p_{3/2}28\ell $\\
 65.27 & 0.008 & 1.23  & 0.040 & $ 2p_{3/2}29\ell $\\
 65.93 & 0.009 & 1.16  & 0.040 & $ 2p_{3/2}30\ell $\\
 66.52 & 0.01  & 1.11  & 0.041 & $ 2p_{3/2}31\ell $\\
\end{tabular}
\caption{\label{tab:strength} Strengths of the individually resolved $2p_jn\ell$ resonances
as obtained from fits of Gaussians to the experimentally observed resonance structures. The
errors given are statistical only (one standard deviation). Systematic uncertainties amount
to less than $\pm 15\%$ for the resonance strengths and less than $0.6\%$ for the resonance
positions.}

\end{table}

\narrowtext

\begin{table}
\begin{tabular}{ld@{}dd@{}d}
 series &  \multicolumn{2}{c}{$2p_{1/2}$}  &   \multicolumn{2}{c}{$2p_{3/2}$} \\
 \hline
 $E_\infty$ (eV), spectroscopic              & 52.95 &            & 74.96 & \\
 $E_\infty$ (eV), this experiment            & 53.19 &$\pm$ 0.01  & 75.34 &$\pm$ 0.01\\
 $\delta$                                    & 0.031 &$\pm$ 0.005 & 0.030 &$\pm$ 0.001\\
 $A_{\rm a}$ (10$^{15}$~s$^{-1}$)            & 0.4 &              & 1.2   &\\
 $A_{\rm r}^{\rm (I)}$ (10$^{9}$~s$^{-1}$)   & 2.0 &              & 5.8   &\\
 $A_{\rm r}^{\rm (II)}$ (10$^{13}$~s$^{-1}$) & 4.1 &$\pm$ 0.6     & 6.7   &$\pm$ 0.5 \\
 $S_0$ (10$^{-27}$~eV$^2$~cm$^2$~s)          & 1.19 &$\pm$ 0.09   & 1.12  &$\pm$ 0.03 \\
 $\ell_{\rm max}$                            & 14.5 &$\pm$ 0.6    & 9.5   &$\pm$ 0.2 \\
\end{tabular}
\caption{Parameters obtained from fits of DR resonance positions and strengths. The series
limits $E_\infty$, their uncertainties and quantum defects $\delta$ result from a fit of the
experimental resonance positions to the Rydberg formula Eq.~(\protect\ref{eq:ryd}). The
total experimental uncertainty of $E_\infty$ is of the order of $\pm 0.5$~eV. The
spectroscopic values listed for comparison are taken from Ref.~\protect\cite{Hin89}. In the
fit of Eq.~(\protect\ref{eq:sigmamodel}) to the measured DR resonance strengths the core
radiative rates $A_{\rm r}^{\rm (I)}$ and the Auger rates $A_{\rm a}$ (as listed) have been
taken from atomic structure calculations\protect\cite{Cow81}. $S_0$ and $A_{\rm r}^{\rm
(II)}$ result from the fit. $\ell_{\rm max}$ has been calculated from
$S_0$.\label{tab:slim}}
\end{table}

\narrowtext

\begin{table}
\begin{tabular}{ldd}
 $i$ & $c_i$ & $E_i$ \\
 &($10^{-2}$cm$^3$ s$^{-1}$ K$^{3/2}$) & (eV) \\
 \hline
  1& 0.417 & 2.91\\
  2& 0.683 & 5.60\\
  3& 1.483 & 19.10\\
  4& 3.184 & 44.54\\
  5& 2.928 & 71.97\\
\end{tabular}
\caption{Ni$^{25+}$ $\Delta n= 0$ plasma DR rate coefficient fit
parameters $c_i$ and $E_i$ according to
Eq.~(\protect\ref{eq:alphafit}). The fit to the full line in
Fig.~\protect\ref{fig:plasma} is accurate to better than 0.5$\%$
for 1.0~eV~$\leq k_{B}T_{e}$. The total uncertainty in the rate
coefficient is $20\%$. \label{tab:tfit}}
\end{table}

\begin{table}
\begin{tabular}{lldrrr}
Ion & Reference & Ion beam energy & \multicolumn{2}{c}{$n_{\rm min}$} & $n_{\rm max}$\\
  & & [Mev/u] & $2p_{1/2}$ & $2p_{3/2}$ & \\
  \hline
  Cl$^{14+}$ & \protect\cite{BSM99} & 7.1 & 23 & 18 & 79 \\
  Ti$^{19+}$ & \protect\cite {bar99}& 4.6 & 27 & 27 & 115\\
  Ni$^{25+}$ & [this work] &5.9 & 31 & 31 & 150\\
\end{tabular}
\caption{Parameters in DRF experiments with Li-like ions. For each ion the field effect has
been quantified by considering an integrated DR rate coefficient which comprises the
resonance strengths of $2p_j n\ell$ DR resonances with $n_{\rm min}\leq n \leq n_{\rm max}$
where $n_{\rm max}$ is the approximate cut-off quantum number due to field ionization in the
charge analyzing dipole magnet. It depends on the ion beam energy and the ion's charge
state\protect\cite{MW97}.\label{tab:allions}}
\end{table}

\begin{figure}
\epsfxsize=8.5cm \centerline{\epsfbox{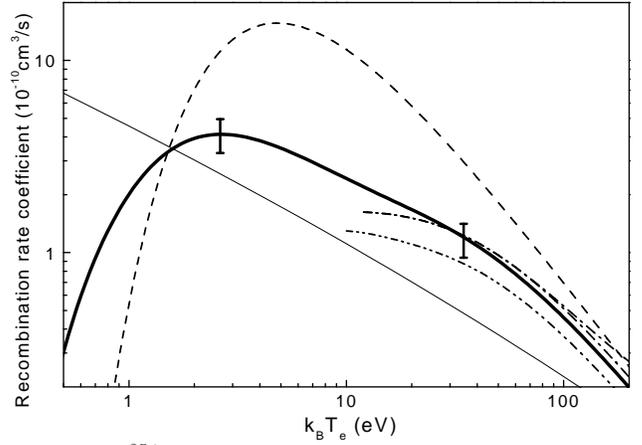}}

\caption[]{Ni$^{25+}$ $\Delta n=0$ DR plasma rate coefficient as derived from our
measurement (thick full line, estimated error $\pm 20\%$ as indicated). Also shown are
theoretical results of Mewe et al.~\protect\cite{MSS80} (dashed line), Romanik
\protect\cite{rom88} (dash-dotted line) and Teng et al.\protect\cite{TSZ94} (dash-dot-dotted
line). At temperatures $k_{\rm B}T_{\rm e}
> 100$~eV two DR rates by Romanik are shown, with the upper one additionally containing
$\Delta n=1$ DR contributions. The RR rate coefficient (thin full
line) has been calculated from the RR cross section given by
Eq.~(\protect\ref{eq:sigmarr}) with $q=25$ and $n_{\rm max}=150$.
\label{fig:plasma}}
\end{figure}

\begin{references}

\bibitem{DV80} J. Dubau and S. Volont\'e, Rep.\ Prog.\ Phys.\ {\bf
43}, 199 (1980).

\bibitem{Mue95} A. M\"{u}ller, in {\it Atomic and Plasma-Material
Interaction Data for Fusion}, Suppl.\ to Nucl.\ Fusion, Vol.\ 6
(IAEA, Vienna, 1995), pp.\ 59--100.

\bibitem{sho69} B. W. Shore, Astrophys.\ J. {\bf 158}, 1205 (1969).

\bibitem{bur64} A. Burgess, Astrophys.\ J. {\bf 139}, 776 (1964).

\bibitem{bur69}
A. Burgess, H. P. Summers, Astrophys.\ J. {\bf 157}, 1007 (1969).

\bibitem{jac76}
V. L. Jacobs, J. Davies, and P. C. Kepple, Phys.\ Rev.\ Lett.\ {\bf
37}, 1390 (1976).

\bibitem{Hahn97}
Y. Hahn, Rep.\ Prog.\ Phys.\ {\bf 60}, 691 (1997).

\bibitem{DS92} P. F. Dittner and S. Datz, in {\it Recombination of
Atomic Ions, NATO ASI Series, Vol.\ 296} edited by W. G. Graham, W.
Fritsch, Y. Hahn, and J. A. Tanis (Plenum Press, New York, 1992) p. 133.

\bibitem{And92} L. H. Andersen, in {\it Recombination of Atomic Ions,
NATO ASI Series, Vol.\ 296} edited by W. G. Graham, W. Fritsch, Y. Hahn,
and J. A. Tanis (Plenum Press, New York, 1992) p.\ 143.

\bibitem{Mueller86/87}
A. M\"uller, D. S. Beli\'c, B. D. DePaola, N. Djuri\'c, G. H. Dunn, D. W.
Mueller, and C. Timmer, Phys.\ Rev.\ Lett.\ {\bf 56} 127 (1986); Phys.\ Rev.\ A
{\bf 36}, 599 (1987).

\bibitem{LNH86} K. LaGattuta, I. Nasser, and Y. Hahn, Phys.\ Rev.\ A {\bf
33}, 2782 (1986).

\bibitem{BGP86} C. Bottcher, D. C. Griffin, and M. S. Pindzola, Phys.\
Rev.\ A {\bf 34}, 860 (1986).

\bibitem{Young94}
A. R. Young, L. D. Gardner, D. W. Savin, G. P. Lafyatis, A. Chutjian, S.
Bliman, and J. L. Kohl, Phys.\ Rev.\ A {\bf 49}, 357 (1994); D. W. Savin, L.
D. Gardner, D. B. Reisenfeld, A. R. Young, and J. L. Kohl, Phys.\ Rev.\ A
{\bf 53}, 280 (1996).

\bibitem{bar97}
T. Bartsch, A. M\"{u}ller, W. Spies, J. Linkemann, H. Danared, D. R. DeWitt,
H. Gao, W. Zong, R. Schuch, A. Wolf, G. H. Dunn, M. S. Pindzola, and D. C.
Griffin, Phys.\ Rev.\ Lett.\ {\bf 79}, 2233 (1997).

\bibitem{rob97}
F. Robicheaux and M. S. Pindzola, Phys.\ Rev.\ Lett.\ {\bf 79}, 2237
(1997).

\bibitem{gri98}
D. C. Griffin, F. Robicheaux, and M.S. Pindzola, Phys.\ Rev.\ A {\bf 57}, 2708 (1998).

\bibitem{rob98}
F. Robicheaux, M. S. Pindzola, and D. C. Griffin, Phys.\ Rev.\ Lett.\
{\bf 80}, 1402 (1998).

\bibitem{HB80}
W. A. Huber and C. Bottcher, J. Phys.\ B {\bf 13}, L399 (1980).

\bibitem{BSM99}
T. Bartsch, S. Schippers, A. M\"{u}ller, C. Brandau, G. Gwinner, A. A.
Saghiri, M. Beutelspacher, M. Grieser, D. Schwalm, A. Wolf, H. Danared,
and G. H. Dunn, Phys.\ Rev.\ Lett.\ {\bf 82}, 3779 (1999).

\bibitem{bar99}
T. Bartsch et al., to be published; T. Bartsch, Dissertation, University of
Giessen (1999).

\bibitem{KKG99} V. Klimenko, L. Ko, and T. F. Gallagher,
Phys.\ Rev.\ Lett.\ {\bf 83}, 3808 (1999).

\bibitem{gri87} D. C. Griffin and M. S. Pindzola, Phys.\ Rev.\ A {\bf 35},
2821 (1987).

\bibitem{BBD98} C. Brandau, F. Bosch, G. Dunn, B. Franzke, A. Hoff\-knecht,
C. Kozhuharov, P. H. Mokler, A. M\"{u}ller, F. Nolden, S. Schippers,
Z. Stachura, M. Steck, T. St\"{o}hlker, T. Wink\-ler, and A. Wolf,
Hyperf.\ Int.\ {\bf 114}, 45 (1998).

\bibitem{JKA89} E. Jaeschke, D. Kr\"{a}mer, W. Arnold, G. Bisoffi, M. Blum, A. Friedrich,
        C. Geyer, M. Grieser, D. Habs, H. W. Heyng, B. Holzer, R. Ihde, M. Jung,
        K. Matl, R. Neumann, A. Noda, W. Ott, B. Povh, R. Repnow,
        F. Schmitt, M. Steck, and E. Steffens, in {\it Proceedings of the
        European Particle Accelerator Conference, Rome, 1988}
        ed.\ by S. Tazzari (World Scientific, Singapore, 1989) p.\ 365.

\bibitem{MW97b} A. M\"{u}ller and A. Wolf, in {\it Accelerator-Based Atomic
Physics Techniques and Applications}, ed.\ by J. C. Austin and S. M.
Shafroth (AIP Press, Woodbury, New York, 1997) p.\ 147.

\bibitem{mue99} A. M\"{u}ller, Phil.\ Trans.\ R.\ Soc.\ Lond.\ A {\bf 357},
1279 (1999).

\bibitem{schi99} S. Schippers, Phys.\ Scr.\ {\bf T80}, 158 (1999).

\bibitem{WGL99} A. Wolf, G. Gwinner, J. Linkemann, A. A. Saghiri, M.
Schmitt, D. Schwalm, M. Grieser, M. Beutelspacher, T. Bartsch, C. Brandau,
A. Hoffknecht, A. M\"{u}ller, S. Schippers, O. Uwira, and D. W. Savin, Nucl.\
Instrum.\ Meth.\ Phys.\ Res.\ A, in print.

\bibitem{Kil92}
G. Kilgus, D. Habs, D. Schwalm, A. Wolf, N. R. Badnell, and A.
M\"{u}ller, Phys.\ Rev.\ A {\bf 46}, 5730 (1992).

\bibitem{Lam96} A. Lampert, A. Wolf, D. Habs, J. Kenntner, G. Kilgus, D. Schwalm,
        M. S. Pindzola, and N. R. Badnell, Phys.\ Rev.\  {\bf 53}, 1413
        (1996).

\bibitem{Gri90}
   M. Grieser, M. Blum, D. Habs, R. V. Hahn, B. Hochadel, E. Jaeschke, C. M. Kleffner,
   M. Stampfer, M. Steck, and A. Noda, in {\it Cooler Rings and their Applications},
   ed.\ by T.  Katayama and A. Noda (World Scientific, Singapore, 1991)
   p.\ 190.

\bibitem{Pas96} S. Pastuszka, U. Schramm, M. Grieser, C. Broude, R. Grimm, D. Habs,
        J. Kenntner, H.-J. Miesner, T. Sch\"{u}{\ss}ler, D. Schwalm, and A. Wolf,
        Nucl.\ Instrum.\ Meth.\ A {\bf 369}, 11 (1996).

\bibitem{Hoc94} B. Hochadel, F. Albrecht, M. Grieser, D. Habs, D. Schwalm, E. Szmola and
        A. Wolf, Nucl.\ Instrum.\ Meth.\ A {\bf 343}, 401 (1994).

\bibitem{Mie96} G. Miersch, D. Habs, J. Kenntner, D. Schwalm, and A. Wolf,
Nucl.\ Instrum.\ Meth.\ A {\bf 369}, 277 (1996).

\bibitem{BS57} H. A. Bethe and E. E. Salpeter, {\it Quantum Mechanics of
One- and Two-Electron Atoms} (Springer-Verlag, Berlin, 1957).

\bibitem{AB90} L. H. Andersen and J. Bolko, Phys.\ Rev.\ A {\bf 42}, 1184 (1990).

\bibitem{Hin89}
H. Hinnov and the TFTR Operating Team, B. Denne and the JET Operating Team,
Phys.\ Rev.\ A {\bf 40}, 4357 (1989).

\bibitem{MW97} A. M\"{u}ller and A. Wolf, Hyperf.\ Int.\ {\bf 107}, 233
(1997).

\bibitem{WKM99} J.-G. Wang, T. Kato, and I. Murakami, Phys.\ Rev.\ A {\bf 60},
2104 (1999).

\bibitem{Cow81} R. D. Cowan, {\it Theory of Atomic Structure and
Spectra} (University of California Press, Berkeley, 1981).

\bibitem{bur65} A. Burgess, Astrophys.\ J.\ {\bf 141}, 1588 (1965).

\bibitem{MSS80} R. Mewe, J. Schrijver, and J. Sylwester, Astron.\ Astrophys.\ {\bf 87}, 55
(1980).

\bibitem{rom88} C. J. Romanik, Astrophys.\ J.\ {\bf 330}, 1022 (1988). An
obvious misprint in the Ni{\scriptsize XXVI} entry of Table 3 is $a_6 =
7.93\times 10^{-9}$. It should read $a_6=7.93\times 10^{-10}$. This value
has been used to plot the curves in Fig.~\ref{fig:plasma} of the present
paper.

\bibitem{TSZ94} H. Teng, B. Sheng, W. Zhang, and Z. Xu, Phys.\ Scr.\
{\bf 49}, 463 (1994).

\bibitem{Che91} M. H. Chen, Phys.\ Rev.\ A {\bf 44}, 4215 (1991).

\bibitem{SBC97}  D. W. Savin, T. Bartsch, M. H. Chen, S. M. Kahn, D. A. Liedahl,
J. Linkemann, A. M\"{u}ller, S. Schippers, M. Schmitt, D. Schwalm, and A.
Wolf, Astrophys.\ J. (Lett.) {\bf 489}, L115 (1997).

\bibitem{NH87} I. Nasser and Y. Hahn, Phys.\ Rev.\ A {\bf 36}, 4704 (1987).

\bibitem{gri98b} D. C. Griffin, D. Mitnik, M. S. Pindzola, and F. Robicheaux, Phys.\ Rev. A
{\bf 58}, 4548 (1998).
\end{references}
\end{document}